\documentclass[10pt,conference]{IEEEtran}
\usepackage[left=0.67in,right=0.67in,top=0.71in,bottom=1.1in]{geometry}
\IEEEoverridecommandlockouts
\usepackage{cite, balance}
\usepackage{amsmath,amssymb,amsfonts}
\usepackage{enumitem}
\usepackage{graphicx}
\usepackage{textcomp}
\usepackage{caption}
\usepackage{subcaption}
\usepackage{dblfloatfix}
\usepackage[utf8]{inputenc}
\usepackage{url}
\usepackage{xcolor}
\usepackage{color}
\usepackage{multirow}
\usepackage{url}
\usepackage[linesnumbered,ruled,vlined]{algorithm2e}
\def\BibTeX{{\rm B\kern-.05em{\sc i\kern-.025em b}\kern-.08em
    T\kern-.1667em\lower.7ex\hbox{E}\kern-.125emX}}

\definecolor{darkblue}{rgb}{0, 0, 0.5}

\usepackage{todonotes}

\begin{document}

\title{Next-Generation Satellite IoT Networks: \\A HAPS-Enabled Solution to Enhance\\ Optical Data Transfer}


\author{
\IEEEauthorblockN{
Ethan Fettes\IEEEauthorrefmark{1},
Pablo G. Madoery\IEEEauthorrefmark{1},
Halim Yanikomeroglu\IEEEauthorrefmark{1}}
Gunes Karabulut Kurt\IEEEauthorrefmark{1}\IEEEauthorrefmark{6},
Colin Bellinger\IEEEauthorrefmark{2},
Stéphane Martel\IEEEauthorrefmark{4},
Khaled Ahmed\IEEEauthorrefmark{4},
Sameera Siddiqui\IEEEauthorrefmark{3}\\

\vspace*{0.3cm}

\IEEEauthorblockA{\IEEEauthorrefmark{1} Non-Terrestrial Networks (NTN) Lab, Department of Systems and Computer Engineering, Carleton University, Canada}
\IEEEauthorblockA{\IEEEauthorrefmark{6} Poly-Grames Research Center, Department of Electrical Engineering, Polytechnique Montréal, Montréal, Canada}
\IEEEauthorblockA{\IEEEauthorrefmark{2} National Research Council Canada, Canada}
\IEEEauthorblockA{\IEEEauthorrefmark{3} Defence Research and Development Canada, Canada}
\IEEEauthorblockA{\IEEEauthorrefmark{4} Satellite Systems, MDA, Canada}

}

\maketitle
\IEEEpubidadjcol

\begin{abstract}
For decades, satellites have facilitated remote internet of things (IoT) services. However, the recent proliferation of increasingly capable sensors and a surge in the number deployed, has led to a substantial growth in the volume of data that needs to be transmitted via satellites. In response to this growing demand, free space optical communication systems have been proposed, as they allow for the use of large bandwidths of unlicensed spectrum, enabling high data rates. However, optical communications are highly vulnerable to weather-induced disruptions, thereby limiting their high potential. This paper proposes the use of high altitude platform station (HAPS) systems in conjunction with delay-tolerant networking techniques to increase the amount of data that can be transmitted to the ground from satellites when compared to the use of traditional ground station network architectures.
The architectural proposal is evaluated in terms of delivery ratio and buffer occupancy, and the subsequent discussion analyzes the advantages, challenges and potential areas for future research.
\end{abstract}


\section{Introduction}
Internet of things (IoT) services, delivered through low Earth orbit (LEO) satellites, have become indispensable to significant sectors of the economy \cite{6GIoT}. Key areas such as agriculture, fisheries, forestry and resource extraction, rely on the data provided by remote IoT sensors to operate efficiently \cite{AISandFire}\cite{EOfuture}.

Although the main component of remote IoT operations is the collection of remote sensing data (access link), another frequently overlooked part of these operations is the communication of collected data to the operator and customers (backhaul link). As the device density in the 6G era is predicted to increase substantially \cite{DownlinkBottleneck}, the amount of data that must be transmitted to the ground will also increase significantly. As current RF feeder link architecture have limited data rates, this could result in a data bottleneck in the system. Data bottlenecks can lead to wasted data collections and is an obstacle in terms of providing higher quality services to customers.

In response to this data bottleneck there has been considerable interest in the use of free space optical (FSO) communications to increase the capacity of the backhaul link in satellite IoT systems \cite{DownlinkBottleneck}. FSO communications provide several benefits including large available bandwidth and are generally unaffected by interference \cite{FSOcean}. This means that FSO communications can provide significantly higher data rates than radio frequency (RF) communications \cite{FSOcean}. To put this into perspective, many satellites communicate with ground stations (GS) using RF over X-band, achieving data rates in the order of gigabits per second at most. FSO communications are capable of data rates on the order of ten's of gigabits per second (Gbps) or terabits per second \cite{RFvsFSO}.

However, there are still numerous challenges that prevent the widespread adoption of high data rate FSO communications. One of them is the greater sensitivity to adverse weather conditions \cite{gsNetOptimization}. Unlike RF communications which are only minimally affected by weather impairments, with the exception of precipitation, FSO communications are impacted by cloud cover \cite{FSOcean}, fog and even dust \cite{DustPaper}. Hence, optical ground stations (OGSs) for remote sensing missions that utilize FSO communications can have significantly lower availability than their RF based counterparts. Traditionally, to reduce the impact of weather conditions on satellite communications, system designers have implemented site diversity \cite{gsNetOptimization}. However, due to the greater risk of weather related disruptions for FSO systems, achieving availability requirements by using site diversity requires a high number of OGSs, increasing costs.

\begin{figure*}[t]
    \centering
    \begin{subfigure}[b]{0.47\textwidth}
        \includegraphics[width=1\linewidth]{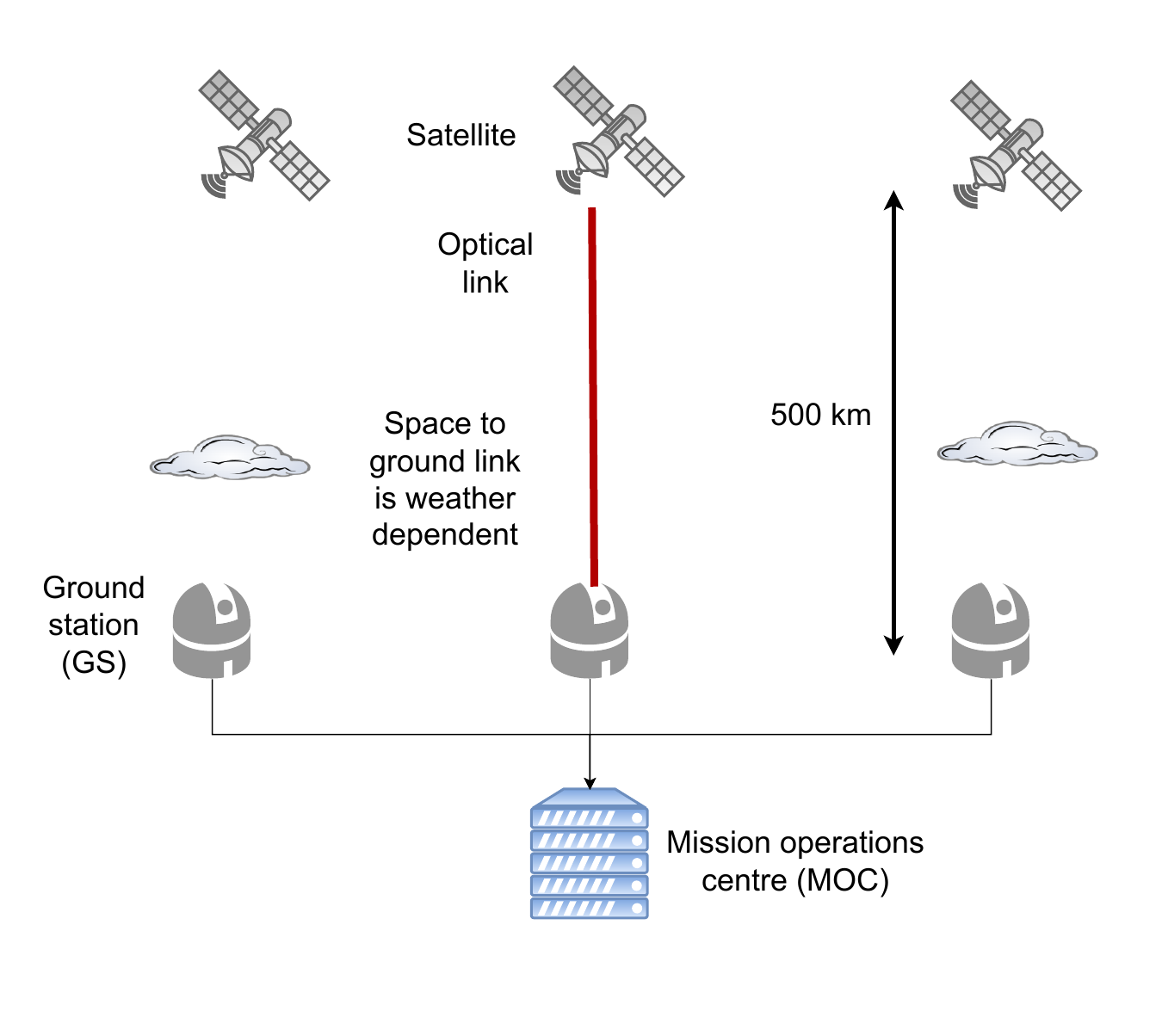}
        \caption{Baseline FSO ground segment architecture}
        \label{fig:baseline}
    \end{subfigure}
    \hfill
    \begin{subfigure}[b]{0.47\textwidth}
        \includegraphics[width=1.0\linewidth]{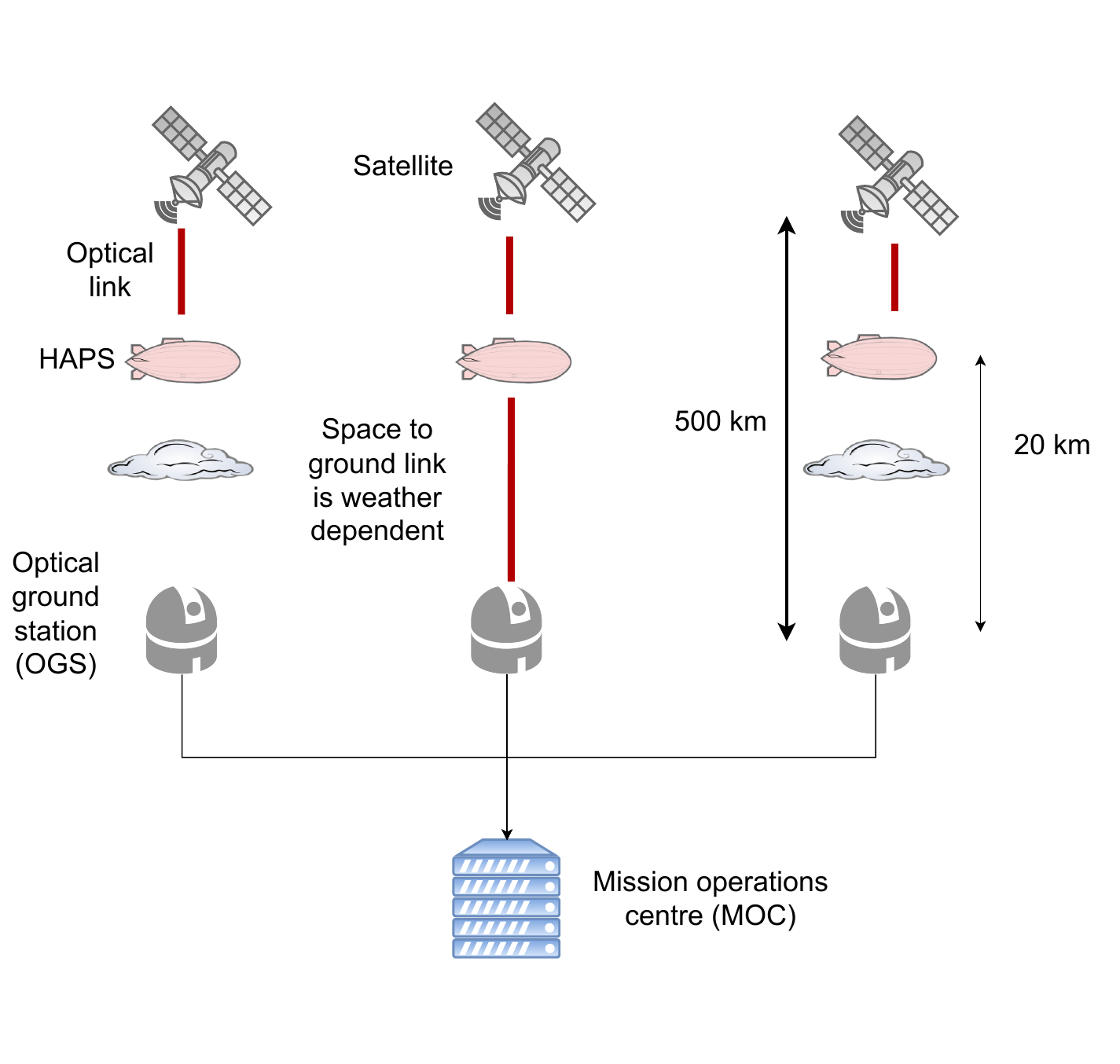}
        \caption{Proposed HAPS-enabled architecture}
        \label{fig:proposed}
    \end{subfigure}

    \caption{Comparison of the baseline and proposed systems architectures}
    \label{fig:architectures}
\end{figure*}

In this paper, we present an architecture harnessing high data rate FSO links, coupled with high altitude platform station (HAPS) systems positioned to buffer data in areas prone to weather disruptions. This innovative approach not only diminishes reliance on OGS numbers but also effectively addresses and mitigates the backhauling bottlenecks encountered in remote IoT satellite applications. Our contributions are as follows:

\begin{itemize}
    \item We propose a novel use case for a HAPS-enabled architecture that leverages FSO links and HAPS to alleviate the backhauling bottleneck in satellite IoT networks in the presence of adverse weather conditions.
    \item We evaluate and compare the proposed architecture with traditional schemes in terms of delivery ratio and buffer occupancy.
    \item We put forth various avenues for future research along with the associated challenges that require resolution.
\end{itemize}

The rest of the paper is structured as follows. Section \ref{sec:background} provides and overview of previous work in this domain,  followed by Section \ref{sec:systemmodel} which outlines the system model. Section \ref{sec:evaluation} delves into the simulation setup and results. Lastly, Section \ref{sec:conclusion} discusses important takeaways from the previous sections and outlines potential avenues for future research.

\section{Background and Motivation}
\label{sec:background}

\begin{figure*}[!b]
\begin{subfigure}{0.5\textwidth}
  \centering
\includegraphics[width=1\linewidth]{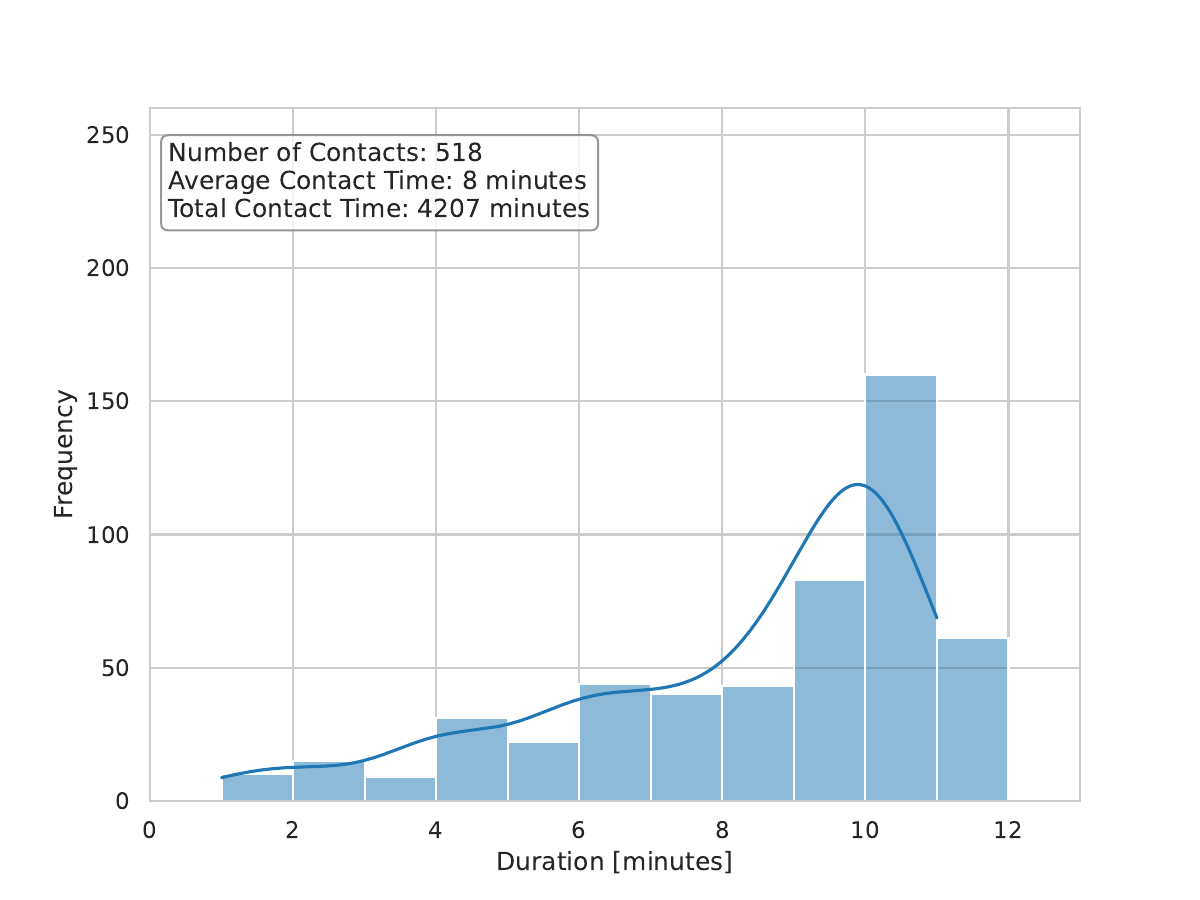} 
  \caption{Histogram of LEO-OGS contact durations.}
  \label{fig:contact_ground}
\end{subfigure}
\hfill
\begin{subfigure}{0.5\textwidth}
  \centering
  \includegraphics[width=1\linewidth]{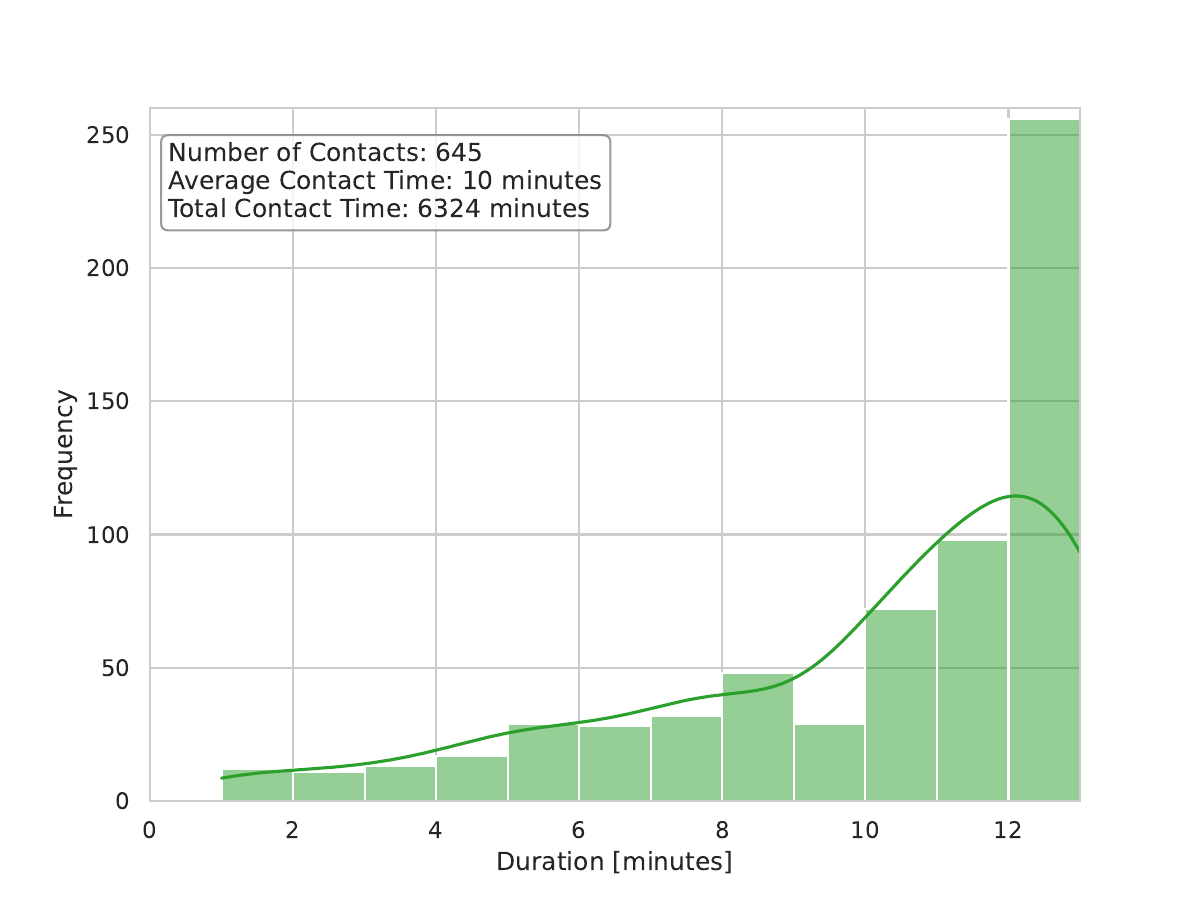} 
  \caption{Histogram of LEO-HAPS contact durations.}
  \label{fig:contact_haps}
\end{subfigure}
\caption{Comparison between LEO-OGS and LEO-HAPS contacts over a 90 day period.}
\label{fig:contact_time}
\end{figure*}

FSO communications are slated to become integral in addressing the rapid increase in data produced by remote IoT sensors. However, they are vulnerable to several atmospheric effects. As the operating wavelengths of FSO devices  are smaller than water droplets contained in clouds and fog, the attenuation experienced due to these weather events is significant \cite{CloudModelPaper}. To address these disruptions site diversity has been the preferred technique. Site diversity involves using many ground station sites to allow the use of an alternate ground station when harsh weather is present. For FSO systems there have been several notable works as discussed below.

In \cite{KoreanLinkAvail}, the availability of OGSs placed in five different South Korean cities is analyzed for a proposed FSO satellite communication system. The authors made use of European Center for Medium-Range Weather Forecast data to inform an atmospheric loss model for space to ground FSO links. Based upon their proposed model the authors subsequently analyze the link reliability for different site diversity solutions.

The number of OGSs and their geographic location has a considerable impact on the availability of the ground segment to the space segment. In \cite{gsNetOptimization}, the authors develop a solution that optimizes the placement of OGSs to achieve a certain availability considering cloud blockage. In this context, care must be taken to avoid placing OGSs in locations where the weather is highly correlated as this introduces increased cost to the system with little increase in availability. Although all OGSs are at least occasionally impeded by adverse weather conditions, due to most weather conditions being confined to the troposphere (0 - 15 Km), assets placed in the stratosphere (15 - 50 Km) can establish communications with satellites largely without these disruptions \cite{SAGINweather}.

HAPS are air assets intended to operate in the stratosphere to provide a variety of services to the ground \cite{SAGnet}. HAPS have been proposed to be a part of vertical heterogeneous networks (VHetNet) \cite{Vhetnetname, SAGsecurity} in conjunction with LEO satellites and ground infrastructure to provide telecommunication services \cite{HAPScloud}. Unlike non-geostationary orbit (NGSO) satellites, HAPS remain stationary or quasi-stationary with respect to the ground. This allows them to have constant line of sight (LOS) with the ground users\cite{HAPScloud}. HAPS are expected to have extensive payload capabilities, allowing for significant computational and storage capacity when compared to other nodes. To take full advantage of HAPS, storage capacity in the context of relaying data from space to ground, a delay tolerant networking protocol, which operate in the presence of intermittent loss of communications, is highly advantageous.

Significant work has been conducted in recent years to develop effective and scalable delay tolerant protocols with a notable use case being deep space networks \cite{CGRTutorial}. Contact Graph Routing (CGR), a delay tolerant networking protocol, uses a bundling layer to allow for the transmission of data between nodes with intermittent or no end-to-end connectivity. This capability is not provided in delay sensitive protocols such as Internet protocols, which requires frequent acknowledgements between source and destination \cite{CGRTutorial}.

\section{System Model and Problem Formulation}
\label{sec:systemmodel}

\subsection{System Architectures}
Traditionally, satellites collect data from remote IoT sensors, and download the data to the ground through distributed ground station networks \cite{AISandFire}. 

Fig. \ref{fig:architectures} depicts a comparison between two system architectures facilitating the download of IoT data to a central mission operations centre (MOC). On the left side, Fig. \ref{fig:baseline} shows a traditional architecture which is presented as a baseline. In this setup, LEO satellites, that have already collected IoT data, utilize FSO links to communicate directly with one or more OGSs distributed globally. Since FSO links are susceptible to adverse weather conditions, any disruption in the link prompts the satellite to store the data in its buffer. The data remains in the buffer until a new contact can be established with another OGS. It is noteworthy that, in this architectural framework, the data download capability is tightly coupled with the prevailing weather conditions at different OGSs.

On the other hand, Fig. \ref{fig:proposed} shows an architecture capable of decoupling the final data download from particular weather conditions. Equipped with buffers, HAPS systems are located between the LEO satellites and the OGS.
In every possible contact between a LEO and a HAPS, the LEO is able to establish a link given that it will not be affected by tropospheric weather conditions. Then, if the HAPS-OGS link is available, the data is sent immediately to the OGS. Otherwise, if the HAPS-OGS link is temporarily unavailable, the data remains in the HAPS buffer until the adverse weather clears.

Both presented architectures exclusively rely on FSO communications, as a combined FSO/RF system would result in higher cost and complexity of the system.

\subsection{Problem Formulation}
The amount of data that can be transmitted from a NGSO satellite to the ground is dependent on several factors. In this paper, we mainly focus in the LOS access and the effect of adverse weather conditions.

\paragraph*{Line of Sight Access}

One of the primary challenges within the baseline architecture stems from the inherent mobility of LEO satellites. The utilization of contact time between LEO and OGS is constrained to the duration of LOS. In contrast, the proposed architecture addresses this limitation by allowing data, upon reaching the HAPS, to be transmitted to the OGS at the achievable data rate, even in instances where the LEO has lost connectivity with the HAPS.

Another factor is the length and frequency of LOS contacts which are determined by both satellite orbit and OGS altitude. Assuming no minimum elevation angle, the length and frequency of LOS contacts can significantly increase for an asset such as a HAPS. 

Fig. \ref{fig:contact_ground} illustrates a histogram with the contact durations between a LEO orbiting at an altitude of 500 km and a single OGS in Ottawa. In contrast, Fig. \ref{fig:contact_haps} shows the contacts between the LEO and a HAPS system located above Ottawa at a height of 20 km. Across a 90 day period, it is observed that the proposed HAPS architecture yields 25\% more contacts and contact durations averaging 25\% longer. Ultimately, this can result in a substantial increase in downloaded data to ground.

\paragraph*{Adverse Weather Conditions}

The presence of adverse weather affects the reliability of the space to ground or air to ground communication link. One method of modelling the reliability of a communication link is in terms of mean time to failure (TTF) and time to recover (TTR). Mean TTF refers to the average amount of time before a failure occurs and TTR is the average amount of time required to recover from this failure. A more reliable link would have a very large TTF and a low TTR, resulting in a high availability. The greater the availability of the ground station the more data can be transmitted to the ground from the space segment. 

As the mean TTF and TTR for a given ground station site is site specific, historical cloud and fog data was used to calculate both values given different cloud thresholds. For both sites of interest Calgary and Ottawa, hourly weather from 2017-2022 was used to calculate the mean TTF and TTR. This data was obtained from both meteoblue and Environment Canada to obtain both cloud and fog hourly data values. Both the TTR and TTF values were defined as separate exponential random variables. First, the rate parameter lambda is calculated as the inverse of the mean TTF or TTR. The equation shown below displays the value for calculating TTF.
\begin{equation}
\lambda=1/mean \textsubscript{TTF}.
\end{equation}
The rate parameter is then used to define the exponential random variable defined below:
\begin{equation}
f\textsubscript{TTF}(x) = 
\begin{cases}
\lambda e^{-\lambda x} & \text{if } x \geq 0 \\
0 & \text{if } x < 0
\end{cases}
.
\end{equation}
These variables are subsequently sampled to obtain TTF and TTR values for simulation purposes.

The cloud threshold for the purposes of this work is a parameter which defines what percentage of clouds in the historical weather data, results in a failure. Hence, a 0\% cloud threshold results in any clouds being present resulting in link failure, while 100\% cloud threshold simulates an OGS that is not affected by cloud cover. In terms of fog data, if fog was present it resulted in a failure condition for all simulations. As the contribution of optical link establishment is small compared to the impact of weather it was disregarded when calculating TTR.

\begin{figure}[]
\begin{subfigure}{0.5\textwidth}
\includegraphics[width=0.98\linewidth]{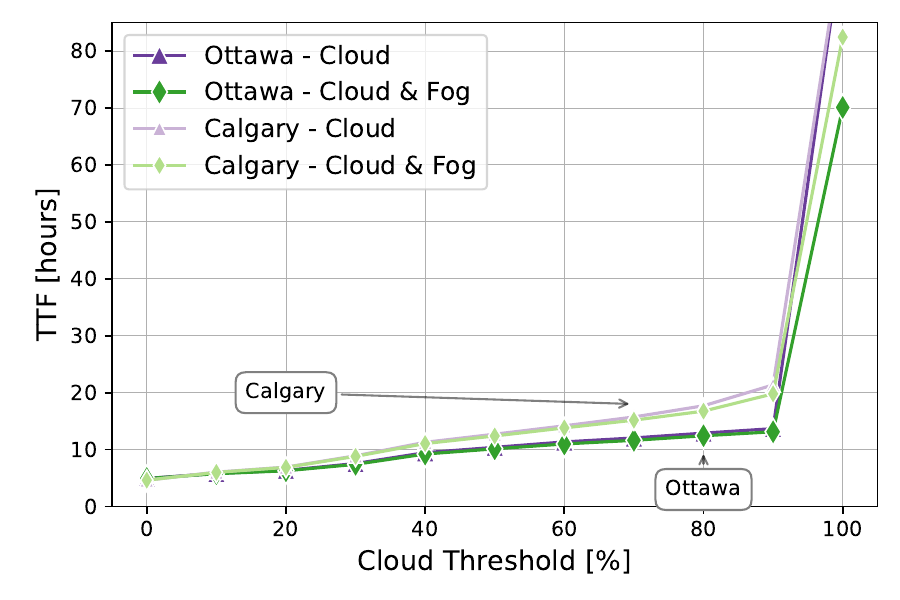} 
  \caption{Time to failure (TTF).}
  \label{fig:TTF}
\end{subfigure}
\begin{subfigure}{0.5\textwidth}
  \includegraphics[width=0.98\linewidth]{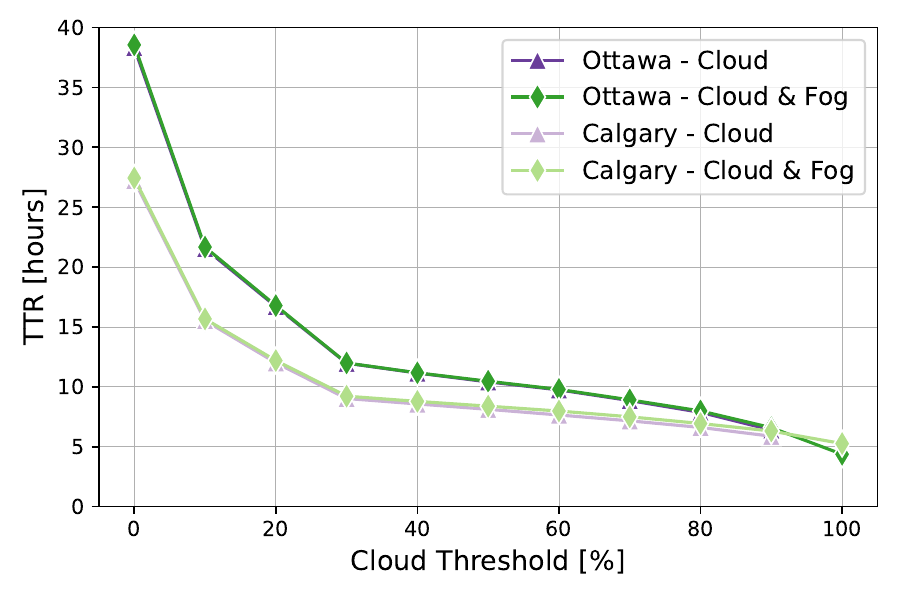} 
  \caption{Time to recover (TTR).}
  \label{fig:TTR}
\end{subfigure}

\caption{Mean time to failure (TTF) and time to recover (TTR) values for Ottawa and Calgary versus a variable cloud cover threshold. Calculations are based on cloud cover data provided by \cite{meteoblue} and fog data provided by \cite{EnvCanada}.}
\label{fig:TTF_TTR}
\end{figure}
Fig. \ref{fig:TTF_TTR} shows the results of the mean TTF and TTR calculations for Calgary and Ottawa. These results show that as the cloud threshold increases, the time to failure increases and the time to recover decreases. Hence, as cloud threshold increases the availability of the site increases. Additionally, from the same figures is possible to observe that at both sites the primary driver of weather-induced link disruptions is cloud cover rather than fog except in the case of a threshold of 100 percent (although the contribution of fog to disruptions is site dependent).

\section{Performance Evaluation}
\label{sec:evaluation}

\subsection{Simulation Setup}

For both the baseline and proposed architectures several configurations were simulated using Matlab Satellite Communication Toolbox \cite{Matlab} and DtnSim \cite{Fraire2017DtnSim}, a discrete-event simulator based on Omnet++. In particular, the following configurations were considered:

\begin{itemize}
    \item 1 LEO satellite transmitting IoT data to 1 OGS in Ottawa.
    \item 1 LEO satellite transmitting IoT data to 2 OGS, one in Ottawa and the other in Calgary.
    \item 1 LEO satellite transmitting IoT data to 1 HAPS connected to a OGS in Ottawa.
    \item 1 LEO satellite transmitting IoT data to 2 HAPS which are in turn connected to OGS in Ottawa and Calgary.
\end{itemize}


Regarding orbital dynamics, the satellite has a polar orbit at an altitude of 500km. This results in several contacts per day for each OGS site with contact assumed to start when LOS was achieved. In terms of the weather modelling, mean TTF and TTR values for both Ottawa and Calgary were calculated from the model described in previous sections. The TTF and TTR were applied to inform the failure model of each OGS, while all other nodes were assumed to not experience failures during the simulation. This implies that only LEO-OGS and HAPS-OGS communications experienced periods of unavailability due to weather conditions.

Each simulation was conducted under the assumption that the LEO had already completed the IoT data collection and was in the process of transmitting data to the ground over the course of one week. Hence, all packets were created at the beginning of the simulation, at the LEO, with the MOC as the final destination. This is consistent with the focus on the backhaul link, rather than the access link. The simulations were conducted at the packet level without considering packet fragmentation. All parameters are summarized in Table \ref{tab:simulation_params}. 

\begin{table}[t]
\centering
\resizebox{0.92\columnwidth}{!}{%
\begin{tabular}{|c|c|c|}
\hline
\multirow{2}{*}{}                    & \textbf{Parameter} & \textbf{Value} \\ \cline{2-3} 
                                     & Simulation time    & 7 days         \\ \hline
\multirow{2}{*}{\textbf{Traffic generated}} & Number of packets                                                                & 1000 \\ \cline{2-3} 
                                     & Packet size        & 20 GB          \\ \hline
\multirow{3}{*}{\textbf{LEO}}        & Altitude           & 500 Km         \\ \cline{2-3} 
                                     & Inclination        & 99.5°          \\ \cline{2-3} 
                                     & Eccentricity      & 0°         
                                               \\ \hline
\multirow{2}{*}{\textbf{OGS Ottawa}}  & Latitude           & 45.4247°       \\ \cline{2-3} 
                                     & Longitude          & -75.6950°      \\ \hline
\multirow{2}{*}{\textbf{OGS Calgary}} & Latitude           & 51.0500°       \\ \cline{2-3} 
                                     & Longitude          & -114.0667°     \\ \hline
\textbf{HAPS}                        & Altitude           & 20 Km          \\ \hline
\textbf{Links}                       & Data rate          & 8 Gbps         \\ \hline
\end{tabular}%
}
\caption{Simulation parameters}
\label{tab:simulation_params}
\end{table}

\subsection{Simulation Results}

Given the intermittent connectivity with orbiting satellite and the absence of strict delay quality of service (QoS) metrics the main metric used to evaluate the performance of the four simulation configurations was delivery ratio. Delivery ratio is measured as the number of packets that arrive at the destination vs the number of packets that attempted to be sent.

\begin{figure}[b]
    \centering
    \includegraphics[width=1.00\linewidth]{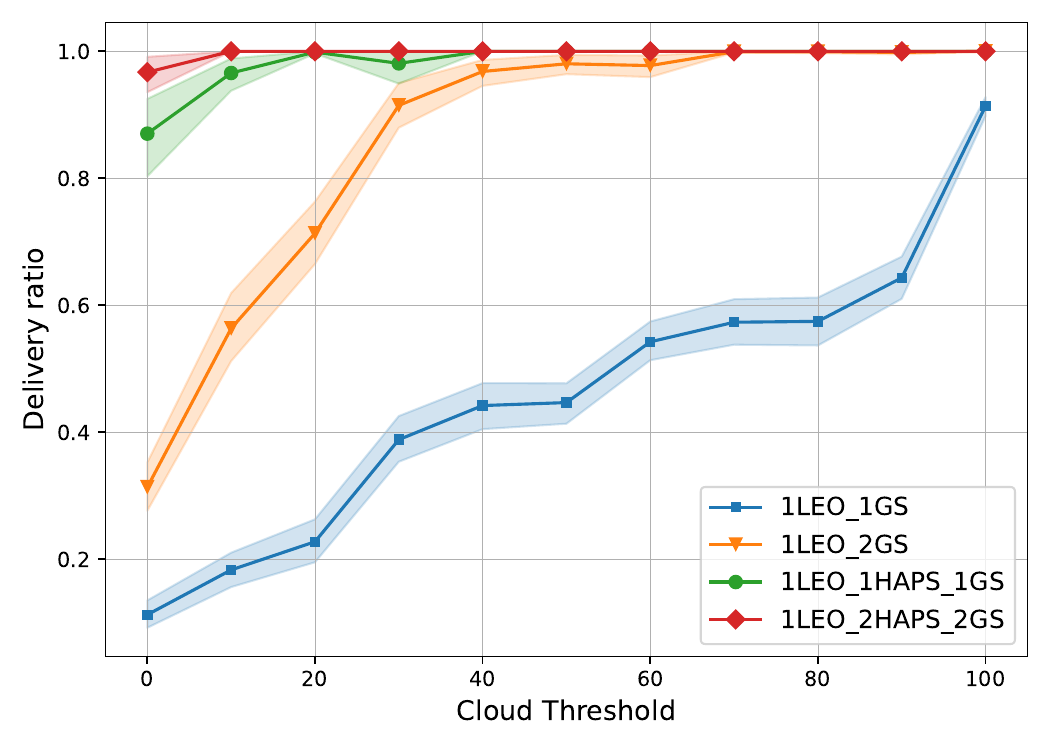}
    \caption{Delivery ratio vs cloud threshold.}
    \label{fig:delivery_ratio}
\end{figure}

There are two main drivers behind using delivery ratio as a metric. Firstly, many IoT use cases require up to date information \cite{AISandFire} hence, getting the data from IoT sensors to the customer must happen promptly. Secondly, as the satellite has a constrained storage capacity, if too much time passes between transmissions, data may need to be discarded or collections be postponed resulting in a loss of data.

\begin{figure}[b]
    \centering
\includegraphics[width=1.1\linewidth]{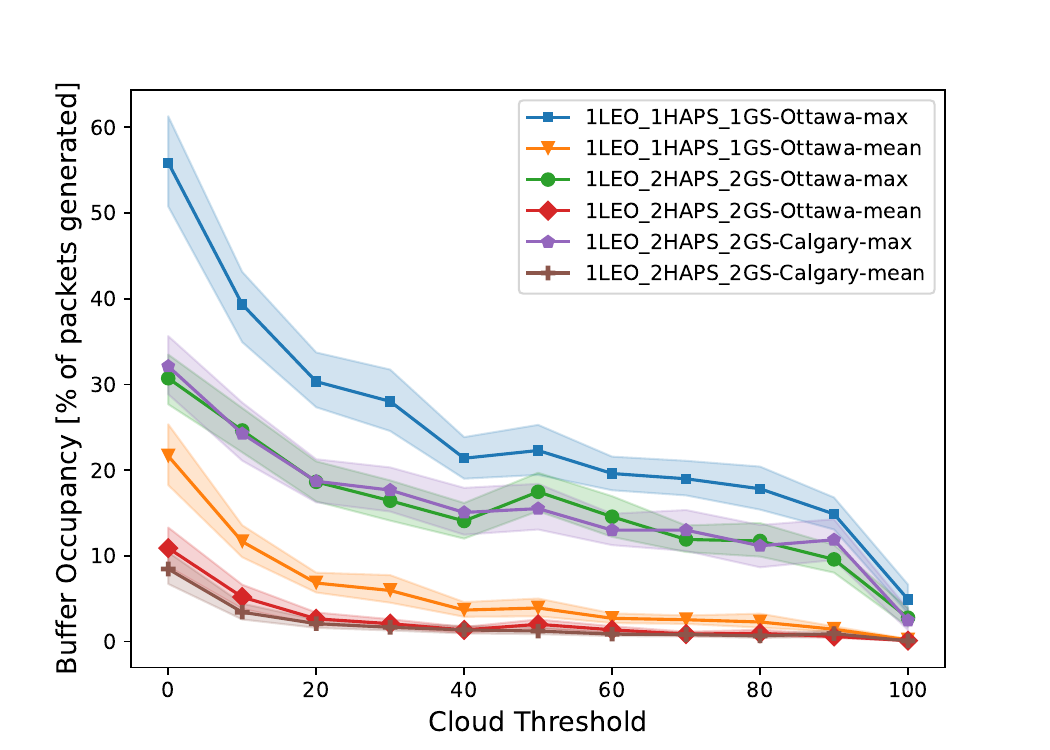}
    \caption{Buffer occupancy of HAPS nodes vs cloud threshold.}
    \label{fig:results_buffer}
\end{figure}

As shown in Fig. \ref{fig:delivery_ratio}, the proposed HAPS-enabled architectures outperform the one and two OGS architectures by large margins of delivery ratio. This is particularly true for low threshold values where we assume that the OGS FSO link is highly sensitive to cloud cover. Furthermore, in order to account for statistical significance, each configuration was run 50 times and averages of the delivery ratio, as well as 95\% confidence intervals, are plotted in the figure.

For the HAPS-enabled architecture configurations, the buffer occupancy of the HAPS nodes was evaluated to determine if there were significant advantages in terms of using two HAPS instead of one. Buffer occupancy percentage, for the purposes of this analysis, is the percentage of the total packets produced at the start of the simulation that are being stored in a network node. As CGR is being utilized, network nodes are able to store packets while waiting for a transmission contact to the next node in the path. The requirements for the buffer capacity of the HAPS nodes are highly relevant for the feasibility of the system.

The results in Fig. \ref{fig:results_buffer} show the results for buffer occupancy vs cloud cover. Each of the curves show how the buffer occupancy mean and max percentage of each HAPS (for both 1 HAPS and 2 HAPS simulation configurations) change with cloud threshold. One can see that with an increase in the number of HAPS there is a decrease in both the maximum and mean buffer occupancy values for all cloud threshold values. In addition, it is observed that as the cloud threshold increases, the buffer occupancy of the HAPS decreases. This is consistent with the expected behavior of the system, as with increasing cloud threshold there is a reduction in times when the HAPS to ground link is impeded by weather disruptions.

Hence, the results presented above demonstrate that HAPS-enabled architectures outperform the baseline architecture in terms of delivery ratio. Furthermore, when employing two sites for the HAPS-enabled architecture configurations, there is a substantial reduction in the required buffer for each HAPS.
\section{Discussion and Future Work}
\label{sec:conclusion}

\subsection{Discussion}
One aspect of the system which warrants further discussion is where data should be stored in the network. 

As a LEO satellite will collect IoT data periodically it requires a certain storage capacity depending on the amount of data produced by sensors and the time between contacts with the ground. Storing large amounts of data on the satellite is both undesirable from a cost perspective and due to space constraints frequently unfeasible. Additionally, if the limited storage capacity of the satellite is exceeded data can be lost. Hence, in the context of the proposed architecture storing data in the HAPS system buffer is highly beneficial when compared to storing data in the space segment. This is due to the ability to upgrade of the HAPS system (unlike satellites) and the high payload capacity for many proposed HAPS systems. However, the buffer capacity of the HAPS system is not infinite and if lower buffer capacity can be used, this would be significant benefits for the feasibility of the proposed architecture by reducing the required payload capacity of the HAPS system.

Although buffer occupancy of HAPS system is an important metric when comparing different network architectures, this must be balanced with the relevancy of satellite buffer occupancy due to high space hardware costs and the inability to upgrade satellite hardware once in orbit.
\subsection{Future Work}
Although the HAPS-based architecture proposed in this paper showed performance advantages over the baseline architecture, there is substantial follow-on work to this study. Firstly, the weather disruption model used in this paper does not capture the partial contacts which may occur during moderately cloudy days, a more advanced model would allow for a more realistic analysis of the effect of cloud cover on FSO communications. Secondly, although CGR was used in the HAPS-based architecture, allowing for storage of data during adverse weather events, a more efficient routing algorithm would take predicted adverse weather conditions into account rather than operating as if they are unpredictable disruptions. Finally, testing this architecture in the context of fronthaul as in \cite{DownlinkBottleneck} would allow for a more comprehensive analysis of the effect of the proposed architecture on a satellite IoT system.

\section*{Acknowledgments}
This work has been supported by the National Research Council Canada's (NRC) High Throughput Secure Networks program within the Optical Satellite Communications Consortium Canada (OSC) framework, MDA and Mitacs. In addition, we would like to thank MeteoBlue and Environment and Climate Change Canada for providing the weather data used in this work.


\end{document}